\documentclass[preprint,11pt]{aastex}
\slugcomment{To appear in Astrophy. J. April 1, 2000 issue, Vol. 532}
\shorttitle{Muench et al.}
\shortauthors{Modeling the Near-Infrared Luminosity Functions}

\begin{document}

\title{Modeling the Near-Infrared Luminosity Functions of \\
 Young Stellar Clusters}

\author{August A. Muench\altaffilmark{1,2} }
\affil{Department of Astronomy, University of Florida, Gainesville, FL 32611 and \\
  Harvard-Smithsonian Center for Astrophysics, Cambridge, MA 02138}
\email{gmuench@cfa.harvard.edu}

\author{Elizabeth A. Lada\altaffilmark{} }
\affil{Department of Astronomy, University of Florida, Gainesville, FL  32611}
\email{lada@astro.ufl.edu}

\and

\author{Charles J. Lada\altaffilmark{} }
\affil{Harvard-Smithsonian Center for Astrophysics, Cambridge, MA 02138}
\email{clada@cfa.harvard.edu}


\altaffiltext{1}{Smithsonian Predoctoral Fellow}
\altaffiltext{2}{present address: Harvard-Smithsonian Center for Astrophysics,
    60 Garden Street, Mail Stop 42, Cambridge, MA 02138 USA}


\begin{abstract}
We present the results of numerical experiments designed to evaluate
the usefulness of near-infrared luminosity functions for constraining
the Initial Mass Function (IMF) of young stellar populations.
We test the sensitivity of the near-infrared K band luminosity 
function (KLF) of a young stellar cluster to variations in the underlying
IMF, star forming history, and pre-main sequence mass-to-luminosity
relations.
Using Monte Carlo techniques, we create a suite of model luminosity 
functions systematically varying each of these basic underlying relations.
From this numerical modeling, we find that 
the luminosity function of a young stellar population is considerably
more sensitive to variations in the underlying initial mass function than
to either variations in the star forming history or assumed 
pre-main-sequence (PMS) mass-to-luminosity relation.
Variations in a cluster's star forming history are also found to
produce significant changes in the KLF.
In particular, we find that the KLFs of young clusters evolve
in a systematic manner with increasing mean age.
Our experiments indicate that variations in the PMS mass-to-luminosity
relation, resulting from differences in adopted PMS tracks
produce only small effects on the form of the model
luminosity functions and that these effects are mostly likely
not detectable observationally.

To illustrate the potential effectiveness of using the KLF of a young
cluster to constrain its IMF, we model the observed K band luminosity
function of the nearby Trapezium cluster.
With knowledge of the star forming history of this cluster obtained
from optical spectroscopic studies, we derive the simplest
underlying IMF whose model luminosity function matches
the observations.
Our derived mass function for the Trapezium spans two
orders of magnitude in stellar mass (5$\;>\;M_\odot\;>\;$0.02) 
and has a peak near the hydrogen burning limit.
Below the hydrogen burning limit, the mass function steadily
decreases with decreasing mass throughout the brown dwarf regime.
Comparison of our IMF with that derived by optical and spectroscopic
methods for the entire Orion Nebula Cluster suggests that modeling 
the KLF is indeed a useful tool for constraining the mass function in young 
stellar clusters particularly at and below the hydrogen burning limit.
\end{abstract}




%

\section{Introduction}
\label{intro}
\setcounter{footnote}{0}

The development of sensitive, large format imaging arrays at near-infrared 
wavelengths has made it possible to obtain statistically significant and 
complete samplings of the near-infrared luminosity functions of very young 
embedded clusters.  
In principle such luminosity functions should provide fundamental constraints 
on the stellar initial mass function (IMF)\footnote{In all cases we will 
refer to the mass function not the mass spectrum of stars. The mass function 
is the number of stars per unit volume per unit {\it log mass}.  
The \cite{sal55} mass function would be a power-law 
$\xi(\:\log(m_{\star})\:)\:=\:m^{\Gamma}$ with a index
$\Gamma\:=\: -1.35$ in this definition.}
 of these clusters and its possible variation in space and time. 
There are several advantages and disadvantages in using young embedded 
clusters to study issues of the initial mass function (for a summary see 
\cite{llm98}; \cite{lad99}). 
The biggest advantage of studying the luminosity functions of very young 
clusters is that the low mass stars in these clusters are brighter than 
at any other time in their evolution. 
Consequently modern NIR detectors even on modest telescopes can completely 
sample the entire IMF of nearby (D $<$ 1 Kpc) embedded clusters down to 
and {\it well below} the hydrogen burning limit. 
However, the youth of the clusters also makes interpreting the luminosity 
functions of these clusters, in terms of an IMF, difficult, since most members 
are still in the pre-main sequence phase of their evolution.
For pre-main sequence stars, there is no unique mass to luminosity relation 
and consequently one must rely on using evolutionary models to understand 
the underlying IMFs of embedded clusters.  

Various groups have modeled the luminosity functions of young clusters using 
realistic stellar mass functions and appropriate mass-luminosity relationships 
(e.g., \cite{zin93, str93, fle94a, ll95, mea96}). 
Zinnecker, McCaughrean \& Wilking (1993) were the first to present model KLFs 
of very young clusters.  
For their model clusters they adopted a ``coeval'' star formation history 
in which all the stars were formed at a single instant of time. 
Moreover, they assumed blackbody radiation to derive bolometric corrections 
and assumed a  single form for the IMF. 
Consequently, their models were not very realistic and they did not 
attempt to fit or directly compare their models to observed KLFs. 
\cite{ll95} (LL95, hereafter) improved on this work by developing 
evolutionary models for the KLFs of young clusters ranging in age from 
10$^6$ to $<$ 10$^7$ yrs, using empirically determined bolometric corrections 
and allowing for non-coeval or continuous star formation in the clusters. 
Moreover, they directly compared their models to observed KLFs of young 
clusters. 
However, similar to the Zinnecker et al. work, Lada \& Lada 
assumed a single underlying IMF (i.e., the \cite{ms79} field star
IMF) and employed the published pre-main sequence evolutionary tracks 
of \cite{dm94}.
In addition, their model KLFs were constructed
for stars having masses between 0.1 and 20 $M_\odot$, since the existing PMS 
tracks did not extend below 0.1 $M_\odot$. Thus their results are only 
valid as long as there are no, or at least very few, brown dwarfs in 
these clusters.
Their modeling efforts produced several interesting results. 
For example, Lada \& Lada found that for a fixed IMF the shape of a cluster 
KLF depends primarily on the duration of star formation and age of the 
cluster and that in general the KLFs broaden with age. 

It is somewhat difficult to evaluate the success of the early modeling work 
to constrain the IMF of embedded clusters because:
1) these early models typically employed one set of PMS tracks with stellar 
masses greater than 0.1 $M_\odot$; 
2) the models only considered one fixed IMF and,
3) the ages and age spreads of the clusters were not well constrained.
Fortunately, improvements in these areas have  recently been made. 
For example, PMS evolutionary tracks have been extended to objects with masses 
well below the hydrogen burning limit. 
In addition, improved age estimates for several clusters such 
as the Trapezium \citep{hil97} and IC 348 \citep{her98} have been made.

Given these advances, we have extended the earlier modeling 
work and have developed a new suite of KLF evolutionary models. 
These new models are based on Monte Carlo techniques and are a 
considerable improvement over the earlier calculations. 
In particular, the new models incorporate improved bolometric corrections 
and new PMS tracks, thereby including objects with masses below 
the hydrogen burning limit.  
In this paper, we describe our new models and present our results 
from this modeling effort.  
In addition, we investigate the utility of using near-infrared luminosity 
functions for interpreting and constraining the IMF of young embedded 
stellar clusters.  
Since the observed luminosity function of a young cluster depends not only 
on the underlying IMF, but also the pre-main sequence mass-to-luminosity 
relation and the star forming history of the clusters, it is important to 
understand how changes in the pre-main sequence mass-to-luminosity relations, 
star forming history and the underlying IMF affect the resulting KLF.  
Therefore, we have systematically varied each of these three basic underlying 
inputs and have evaluated the sensitivity of the KLF to each input.  
Finally, we illustrate the effectiveness of using the KLF of a young cluster 
to constrain its IMF by modeling the KLF of the Trapezium cluster for which 
detailed knowledge of its star formation 
history is known from optical spectroscopic studies \citep{hil97}.

\section{Modeling the Luminosity Function}
\label{models}

The observed luminosity function (LF) for a young cluster depends
on the pre-main sequence mass to luminosity (M-L) relation, the star
forming history (SFH) of the cluster and 
the cluster's underlying initial mass function (IMF).
We constructed model K-band LFs (KLFs) for a suite of synthetic young 
clusters using varying PMS M-L relations, SFHs and IMFs. 
We first specified a functional form for the SFH and the IMF for 
each synthetic cluster.
The ages and masses of the individual cluster members were then sampled from
these adopted functions using a Monte Carlo rejection method algorithm.
We derived the monochromatic magnitudes for each model star from its age
and mass by using theoretical pre-main sequence evolutionary models and
empirical bolometric corrections taken from the literature.
For each synthetic cluster, we binned the resulting magnitudes in 
half magnitude bins to compare to observed cluster luminosity functions.
Our standard cluster contained 1000 stars and for each SFH and IMF
 we produced 100 independent luminosity functions.
We computed the mean luminosity function from these 100 realizations, 
 and report the one sigma standard deviation of the
 computed mean.

For the models presented in this paper, we assumed a constant star
 formation rate over the SFH of the cluster.
We parameterized the SFH using a "mean age", $\tau$, and
 an "age spread", $\Delta\tau$.
For example, a coeval cluster will have no age spread and 
 $\Delta\tau/\tau = 0.0$.
A cluster with the largest possible age spread would have
 $\Delta\tau/\tau = 2.0$ with star formation
 starting $2\times\tau$ years ago and continuing to the present.
Figure \ref{age_agespread} illustrates these definitions.
Our model SFH then approximates any SFH to first order by using
 the most common age of the members and a rough age spread.

In our standard model, stars have masses from 80 to 0.02 M$_\odot$. 
For all the models considered here, the youngest 
($1-5\:\times\:10^{5}$ years) cluster members with masses greater than
 5 $M_\odot$ will have already reached the Zero Age Main Sequence (ZAMS).
For these massive stars, we converted their age and mass 
to bolometric luminosity and effective temperature using 
 a theoretical ZAMS derived from \cite{sch92}.
Between 1 and 5 M$_\odot$, we used pre-main sequence evolutionary models
 from either \cite{ps93} or \cite{ber96a}.
For stars below 1 M$_\odot$ and brown dwarfs, we used PMS evolutionary 
 models from D'Antona \& Mazzitelli (1994,1998).
Appendix \ref{pmsappendix} describes how these three regimes were combined.
Using the bolometric luminosity and effective temperatures derived
 from these stellar models, we converted to an absolute magnitude using the
 formula:
\begin{equation}
M_\lambda = M_{bol,\odot} - 2.5 \:\times\: \log(L/L_{\odot}) - BC_{\lambda}(T_{eff})
\label{converteq}
\end{equation}
Our empirical bolometric corrections were tabulated as functions of
 effective temperature and were taken from the literature.
In addition, we assumed $M_{bol,\odot} = 4.75$.
We list the sources of the bolometric corrections 
 in Appendix \ref{boloappendix}.

For the purposes of this paper, we will only present an analysis of the 
 K-band (2.2 $\mu$m) luminosity functions for our model stars, though any 
 monochromatic luminosity functions could be calculated.
Note that our models did not include the
 effects of excess infrared flux due to unresolved binaries or
 circumstellar accretion disks around PMS stars.
However, by choosing relatively wide (half magnitude) bins for the luminosity
 function, we can minimize the variable effects of infrared excess flux.
The model KLFs for our young clusters did not include the effects of
 differential interstellar extinction common in young star forming regions.
We did not include protostars in our luminosity function models since
 the contribution of proto-stellar objects to the luminosity function
 of most star forming regions should be small \citep{fle94a,fle94b}.
Finally, we defined our input mass function as a single star mass
 function.

\section{Experiments}
\label{exps}

\subsection{Evolutionary Pre-Main Sequence Models}
\label{expml}

The evolution of pre-main sequence (PMS) stars across the HR
 diagram and onto the main sequence is not observationally well constrained.
Details of PMS evolution rely heavily upon theoretical PMS models.
These theoretical PMS models vary in their predictions depending on the
 numerical methods and theoretical assumptions used in their creation.
Since these PMS models are used to convert from a stellar age and mass to
 a monochromatic magnitude, the resulting luminosity functions
 will depend to some degree on the PMS evolutionary models which are chosen.
To evaluate how PMS models with different input physics, chemical 
 abundances or effective mass ranges affect the shape and form of a model
 luminosity function, we constructed and compared model luminosity functions
 calculated assuming different PMS models.

For these experiments, we fixed the initial mass function to have a
 log-normal distribution:
\begin{eqnarray}
 \xi(\log(\frac{M}{M_{\odot}})) & = & 
 c1*\exp(-c2*(\log(\frac{M}{M_{\odot}}) - c3)^2)),
\label{ms79eq}
\end{eqnarray}
 where c1 is a normalization constant, c2 equals $1/(2\log(\sigma)^2)$,
 c3 equals $\overline{log(\frac{M}{M_{\odot}})}$ or the mean log mass of the 
 distribution, and $\sigma$ is the variance of this mean.
The mean and variance that we adopted correspond to the field star mass
 function given by \cite{ms79} (MS79, here after), 
having constants of $\mbox{c2} = 1.09$ and $\mbox{c3} = -1.02$ 
 or a mean mass of 0.0955 M$_{\odot}$.
We produced a suite of model clusters with a range of mean ages from
 0.2 to 15 Myrs and age spreads from coeval to twice the mean age of the 
 model cluster.
For the purposes of evaluating the effects of using different input
 PMS models, we only directly compared KLF models having identical star 
 forming histories.

\subsubsection{\cite{dm94}: Differing Input Physics \label{dm94test}}

\cite{dm94} (hereafter DM94) calculated four different sets of 
 evolutionary PMS models varying two input physical parameters, the 
 input opacity tables and the treatments of internal convection.
Table \ref{pmstable} summarizes the four combinations of input physics
 and other parameters of the DM94 PMS models.
Only one of these data sets contained stars with masses less than the 
 hydrogen burning limit.
Consequently, we used a common range of 
 stellar masses from 2.5 to 0.1 M$_\odot$ to compute different 
 model KLFs using the four sets of DM94 PMS models.
Figure \ref{dm94compare} compares synthetic KLFs computed from the DM94 PMS 
 models for coeval models with mean ages of 1 and 7 million years, respectively.
In Figure \ref{dm94compare}, different symbols correspond to different
 input opacity tables in the PMS models used.
For the 1 million year coeval models, the two KLFs corresponding to PMS 
 models with Kurucz opacities are essentially indistinguishable, 
 indicating that the KLFs are insensitive to the convection model used.
The two model KLFs corresponding to PMS models with Alexander opacities
 exhibit a relatively narrow but significant feature or peak between 
 M$_K$ $\sim$ 3-4 which is not apparent in the KLFs with Kurucz opacities.
The position of this spike is different for the two convection models used
 with the Alexander opacities.
This feature is due to deuterium burning which causes 
 a slowing of the stellar luminosity evolution \citep{zin93} and
 therefore results in a pile up of stars in the luminosity function.
The deuterium burning spike is absent in the 7 Myr coeval model in Figure
 \ref{dm94compare}, and in all coeval models with mean ages greater than 
  2-3 Myrs for stars above the hydrogen burning limit.
The onset of deuterium burning is a function of stellar mass. 
Low mass stars contract more slowly than higher mass stars and begin
 burning deuterium after high mass stars. 
However by 3 Myrs, even stars at the hydrogen burning limit would have
 burned all of their initial deuterium abundance.

A second feature of interest in the KLFs is the spike/dip at M$_K$ = 2 in the
 7 Myr coeval model.
It is present in all four 7 Myr KLFs and in all KLFs with mean ages
 greater than 3-4 Myr. 
This feature is the result of stars reaching a luminosity maximum on
 radiative tracks before beginning hydrogen burning and moving to 
 lower luminosities on the main sequence \citep{ibs65}.
We refer to this feature as the luminosity maximum spike (LMS).
The luminosity maximum spike has been studied by \cite{bel97}
in intermediate age (50-100 Myrs) clusters
and these authors have used it to study the age of the Pleiades open 
cluster \citep{bel98}.

Model KLFs appear degenerate in the absence of the deuterium burning spike.
The existence of a deuterium spike removes the degeneracy and 
 differentiates between the two different PMS opacity models.
Moreover, the position of the deuterium burning spike can differentiate
 between the two convection treatments but only for Alexander opacities.
However, only the youngest clusters exhibit a deuterium burning spike.
Model KLFs computed with different PMS tracks and with mean ages greater 
 then 2-3 Myrs are essentially indistinguishable from each other and 
 consequently insensitive to the input physics of the PMS models.
Note that the deuterium burning spike is most prominent when 
 deuterium burning occurs in those stars with masses at the peak of
 the chosen IMF, which for models discussed here occurs at the
 hydrogen burning limit (mean ages 1-2 Myrs).

Introducing an age spread to the cluster star forming history diminishes the 
 differences between the KLFs for all four DM94 and at any cluster mean age.
While we fully describe the effects of age and age spread on the model KLFs
 in section \ref{expsfh}, these result implies that except in the youngest
 clusters, the KLF will be observationally insensitive to 
 variations in the input physics of the PMS models.

To further study how different PMS models affect the model KLFs,
 we compared the PMS models of DM94 with the more recent and improved
 calculations of \cite{dm98} (DM98).
Table \ref{pmstable} lists the relevant characteristics of the DM98 models.
We constructed model KLFs computed with the standard mass range (0.02 to 80 
 M$_\odot$) using the DM94 ACM and DM98 d2.5 PMS models. 
These two PMS models have similar deuterium abundances 
 but DM98 have advancements to the opacity table and treatment of convection
 as well as a new equation of state.
Figure \ref{dm94vdm98} compares model KLFs using these two PMS models
 with a coeval cluster SFH and mean ages of 0.8 and 5 million years.
In general the overall shapes of the model KLFs from the two different 
 PMS models are quite similar but some minor differences can be quantified.
First, the DM98 model KLFs are somewhat narrower and
 have peaks shifted to slightly brighter magnitudes than those KLFs
 corresponding to the DM94 ACM model.
This was a consistent result for all cluster ages and star forming histories.
Second, the largest differences between the model KLFs occur at the 
 faint end.
This is where DM98 describe the largest differences in their PMS models
 with respect to the DM94 PMS models.
DM98 PMS models have a very different resulting mass-effective temperature 
 relation for low mass stars and brown dwarfs than DM94.
Since the K band bolometric correction is fairly insensitive to 
 effective temperature for stars cooler than 3500K (see Appendix 
 \ref{boloappendix}), these changes do not radically affect the model KLF.
Further, DM98 PMS models have larger luminosities for the low
 mass stars and young brown dwarfs compared to DM94.
Likewise the DM98 model KLFs are shifted
 to brighter magnitudes with respect to DM94 for the faint end of the KLF.
However, these differences in the KLFs are relatively small and it would 
 be difficult, observationally, to distinguish between them.

\subsubsection{\cite{dm98}: Variations in Deuterium Abundance \label{dm97test}}

The DM98 PMS models were specifically created to study the effects of
 varying the initial deuterium abundance for PMS evolutionary calculations.
It is unclear how much deuterium pre-main sequence stars might contain
 as they evolve from the birthline toward the main sequence.
And there is little observational evidence to constrain this parameter,
 so it should be considered as an ambiguity in modeling the KLFs.
We studied the effects of the deuterium abundance on the KLFs
 by experimenting with the three PMS models presented by DM98.
The opacities used by DM98 in their PMS models are advancements to those
 in the DM94 PMS models which produced a deuterium burning spike in the
 KLFs of Figure \ref{dm94compare}.
DM98 input physics and deuterium abundances are
 summarized in Table \ref{pmstable}. 
We produced model KLFs using the three DM98 PMS models, d1.5, d2.5, and d4.5,
 so labeled by their respective deuterium abundance ratios, e.g.,
 model d1.5 equals a deuterium abundance of $1.0\:\times\:10^{-5}$.
Respectively, these three PMS models have deuterium abundances  
 of one half, one and two times the interstellar deuterium 
 abundance, which is $[D/H]_0 \:\: = \:\: 2.0\:\times\:10^{-5}$.
Figure \ref{dm97compare} compares model KLFs derived from these PMS models
 for mean ages of 2 and 7 Myr and both coeval and $\Delta\tau/\tau\:=\:2.0$
 age spreads.
Comparing the coeval models it is clear that increasing the [D/H] 
 abundance shifts the deuterium burning spike to brighter magnitudes and
 increases its size.
The deuterium burning peak disappeared from the d1.5 KLF by 3 Myr, the
 d2.5 KLFs by 10 Myr and from the d4.5 KLFs not until beyond 10 Myr.
For model KLFs shown in Figure \ref{dm97compare} with the maximum age spread, 
 variations in the KLFs due to changes in the initial deuterium abundance 
 are too small to be observable.
The main result here is that variations in the $[D/H]$ ratio only produce
 significant (i.e., observable) differences in the model KLFs of 
 coeval (no age spread) clusters.
For these clusters variations in deuterium abundance affects the location
 and size of the deuterium burning feature and this occurs only in younger
 ($\tau\:\:<\:$ 3 Myr) clusters or for the highest deuterium abundances.
Once stars have undergone deuterium burning, their KLFs are identical.
Again, the presence of an age spread dilutes the deuterium burning
 feature rendering the form of the cluster KLF independent of 
 the $[D/H]$ ratio.

\subsubsection{Effective Mass Ranges for PMS Models\label{massrange}}

We investigated the effects of using different IMF mass ranges by 
 comparing model KLFs with the standard mass range (0.02 to 80 M$_\odot$)
 to model KLFs with a truncated mass range (0.1 to 2.5 M$_\odot$), i.e.,
 one excluding brown dwarfs, intermediate or high mass stars.
This experiment is useful for comparing our model LFs to prior 
 LF modeling by other authors who typically did not include stars 
 below the hydrogen burning limit or did not include high mass stars.
Figure \ref{mltrunc} compares model KLFs with truncated and 
 standard mass ranges for two different star forming histories.
For a coeval SFH (upper panel, mean age 3.0 Myr), a truncation in 
 the mass range produces a truncation in the model KLFs at the highest
 and lowest magnitude bins.
However, with an age spread (lower panel, same mean age, 
 $\Delta\tau/\tau\:=\:2.0$), the truncated model KLF is deficient 
 in stars over a wider range of magnitudes, and the two KLFs 
 are similar only over a narrow range of magnitudes.
The form of the cluster KLF is clearly very sensitive to the adopted mass 
 range of the underlying IMF. 

\subsection{Star Formation History} 
\label{expsfh}

As shown in the experiments of sections \ref{dm94test} and \ref{dm97test},
 mean age and age spread have an important effect on the KLF.
To more fully explore this, we created model KLFs with
 a range of mean ages and age spreads, using a single underlying
 mass function, and a fixed set of PMS tracks.
For these experiments, we used the same IMF as in section \ref{expml}
 (see equation \ref{ms79eq}).
As in the previous section, we considered two mass ranges for the IMF,
 one range with stars down to the 0.1 $M_\odot$ and one including 
 brown dwarfs with masses down to 0.02 $M_\odot$.
We adopted a PMS evolutionary model that is our standard combination 
 of DM98 d2.5 PMS models, \cite{ber96a} intermediate-mass PMS models, 
 and \cite{sch92} ZAMS models (see Appendix \ref{pmsappendix}).
We compared the effects of changing the mean age and age spread by
 studying how model KLFs evolve with time.

Figure \ref{meanages} compares model KLFs with different 
 mean ages and cluster age spreads.
Each panel simultaneously displays a one, three and ten million
 year mean age cluster KLF for a specific $\Delta\tau/\tau$.
For a given age spread, the models clearly shift to fainter magnitudes 
 with increasing mean age.
For small age spreads, the deuterium burning feature also evolves to 
 fainter magnitudes with time appearing at M$_K$ = 3.5 at 1 Myrs and 
 M$_K$ = 5.5 at 3 Myrs, and M$_K$ = 8 at 10 Myrs.
To quantify the KLF evolution with time, we calculated the mean K magnitude of 
 the model KLF at each mean age from 0.5 to 10 Myrs and for a range of 
 age spreads.
In Figure \ref{meank}, we plot the KLF mean magnitude versus the 
 cluster mean age.
Also we show this value for the two limiting values of $\Delta\tau/\tau$.
Two sets of curves are plotted, the upper corresponding to an underlying
 cluster IMF with brown dwarfs (standard IMF) and the lower to an 
 underlying IMF having no stars with masses less than 0.1 M$_\odot$. 

The mean K magnitude of the model KLFs evolves over 2 magnitudes in the
 first 10 Myrs of the cluster lifetime, regardless of the age spread or 
 the mass range over which the IMF was considered.
Age spread has little effect except to slightly shift the KLFs to
 brighter magnitudes.
The evolution of the mean K magnitude proceeds most quickly in the
 first 3 million years where the models evolve by 1 full magnitude.
The model KLFs without brown dwarfs naturally have significantly
 brighter mean values but for these KLFs the mean K magnitude evolves 
 similarly to the standard models.
This indicates that the KLFs are more sensitive to changes in the 
 underlying IMF than to changes in the cluster star forming history.
We also studied the width of the model KLFs and found that KLFs widen
 systematically with time as was shown by LL95.

Variations in the mean cluster age produce more significant changes in the
the model KLFs than do changes in the cluster age spread.
We show in Figure \ref{deltat}, model KLFs for two mean ages and for both of
 these mean ages we show the four different age spreads from 
 Figure \ref{meanages}.
For a given mean age, it would be difficult to observationally 
 distinguish clusters with differing age spreads.
In detail, models with differing age spreads do exhibit differences
 in the prominence of the deuterium burning spike and the maximum
 luminosity dip/spike.
At what point can one distinguish a coeval model KLF
 from a model KLF with an age spread?
To answer this question, we compared model KLFs with increasing 
 age spread to a coeval model of the same mean age.   
Using the $\chi^2$ test, we distinguished the age spread at which
 the models KLFs no longer appear to be coeval.
The general trend from our test is that for an increasing mean age, we 
 require a steadily increasing age spread to distinguish the models 
 from a coeval KLF.
For mean ages up to 5 Myrs, a cluster KLF appears coeval until the
 duration of star formation equals the age of the cluster.
Between 5 and 10 Myr mean ages, the deuterium burning feature is present
 among the brown dwarfs but is not very prominent in the model KLFs. 
Only a very small age spread is required to erase it from the model KLFs and 
 the models no longer appear "coeval" with only a small amount of age spread.
Once the deuterium burning feature is lost from the M-L relation,
 the models require very large and probably unrealistic age spreads for
 them to significantly differ from a coeval model of the same mean age.

\subsection{Initial Mass Function} 
\label{expimf}

We varied the underlying initial mass function of a young cluster
 to test the influence of the input IMF on the model KLFs.
In previous sections we used a single IMF equivalent to the 
 log-normal (MS79) mass function and only changed the mass limits
 to this IMF.
To test the sensitivity of the KLF to variations in the underlying IMF,
 we adopted a simple two power-law IMF characterized by the form:
\begin{eqnarray}
\label{2poweq}
\xi( \log m_\star ) & = & A \: \times \: m_{\star}^{\Gamma} 
\end{eqnarray}
for each power-law segment.
We used the symbol $\Gamma$ to represent the power-law
 index of the mass function. 
Two power-law segments are joined at a mass, $m_{break}$.  
The power-law index for masses greater than some $m_{break}$, 
 is $\Gamma_{1}$. 
For masses below the $m_{break}$ value, the IMF 
 index is $\Gamma_{2}$. 
Figure \ref{massfunc} illustrates and compares a sample of the 
 mass functions we used.
The log-normal IMF shown is that used in sections \ref{expml} and  
 \ref{expsfh} and follows the MS79 parameterization. 
The example two power-law IMFs shown in Figure \ref{massfunc} have 
 $\Gamma_{1} = -1.35$, $m_{break} = 0.5\:M_\odot$ and $\Gamma_{2}$ 
 varying from -1.35 to +1.0. 

In these experiments, we varied $\Gamma_{1}$ values from -2.5 to -0.25, 
 $m_{break}$ from 0.06 to 1.5 $M_\odot$ and $\Gamma_{2}$ values from
 -1.35 to +2.0.  
Figure \ref{2powerklf} displays some of the model KLFs and the corresponding
 underlying IMFs.
The cluster star forming history used for these models has a mean
 age of 5 Myrs and a $\Delta\tau/\tau$ = 1.0, or an age spread of 5 Myrs.
We show model KLFs normalized to the bright end of the KLF where
 the underlying IMF power-law indices have 
 identical $\Gamma_{1}$ slopes equal to -1.35. 
This example uses a $m_{break} = 0.5 M_\odot$ and five $\Gamma_{2}$ values
 equal to -1.35, -0.40, 0.0, +0.40 and +1.35. 
The most steeply rising KLF corresponds to a single Salpeter (1955)
 power-law IMF over the entire standard mass range.

Model KLFs display variations due to changes in all three parameters
 of the two power-law IMF. 
In Figure \ref{2powerklf}, the effects of changing $\Gamma_{2}$ 
 are large and the differences between KLFs with a slightly rising and 
 a slightly falling IMF below the $m_{break}$ are significant. 
Varying the $m_{break}$ produces shifts in the peak of the model KLFs. 
Another result of these tests is that over the range of K magnitudes 
 governed by a single underlying IMF power-law,
 the model KLF tends to be characterized by a power-law like slope.
%
%
%
%
%
 
This is true both for the bright and faint slopes of the model KLFs away
from the turnover caused by the m$_{break}$ in the model IMF.
However, we do note that features (e.g., peaks or dips) can arise in the LF
 of a cluster which do not correspond to features in the underlying Mass Function,
 but instead, result from changes in the slope of the Mass-Luminosity
 relation \citep{maz72,dm83,ktg93}.
Such a feature is present at M$_{K}$ = 5.5 in the model KLFs of
 Figure \ref{2powerklf} and this peak is the result of a change of slope
 in the Mass-Luminosity relation due to deuterium burning.
This small peak is not related to a corresponding feature in the underlying
 Mass Function.
Moreover, it is quite small and is not likely to be observationally detectable
 (see Section \ref{expml}).
In addition, the steep downward trend often seen in the last bin of the model
 KLFs is the result of truncating the mass range for the underlying IMF at
 0.02 M$_\odot$, which is the lower mass limit of the model Mass-Luminosity
 relation.

In summary, these calculations clearly show that the shape of the model
 KLF is very sensitive to variations in the underlying cluster IMF.
Indeed, modest variations in the cluster IMF produce significantly
 greater responses in the model KLFs than do variations in the SFH
 and PMS model input physics.

\section{Discussion}

\subsection{Results and Implications of Numerical Experiments}
\label{discussexp}

From these numerical experiments which evaluate the sensitivity of the
K-band luminosity function to variations in three of its fundamental
physical parameters: its underlying IMF, its star forming history,
and its mass-to-luminosity relation, we find that the KLF of a young
cluster is more sensitive to variations in its underlying IMF than
to either variations in the star forming history or the PMS 
mass-to-luminosity relation.

We also find that variations in the cluster mean age can produce 
a significant response in the KLF of a young cluster. In particular,
we find that the KLF systematically evolves with time. 
Both the mean magnitude and the width of the KLF increase with
increasing mean age, confirming the results of earlier modeling (LL95).
At the same time, variations in the cluster age spread are found
to have a small effect on the form of the KLF and would likely 
be difficult to distinguish observationally.

Except for the youngest and purely coeval clusters, we find that the 
synthetic KLFs appear relatively insensitive to the adopted PMS 
evolutionary models (at least for the range of PMS models considered here).
In the youngest coeval clusters, the location and size of the 
deuterium burning spike in the KLF was found to depend sensitively 
on the PMS tracks adopted for the underlying stars.
However, we find that even a small amount of age spread broadens
the spike and would make it observationally difficult to detect.

We conclude from these experiments that the KLF of a young stellar population
can be used to place interesting constraints on the form of the cluster's
underlying IMF, provided an independent estimate
of the cluster mean age is available.
The most direct method of determining the mean age of a young cluster is to
obtain optical or infrared spectra and place the objects on the HR diagram. 
Through comparison to theoretical PMS tracks, the ages of the stars are 
determined and a mean age for the cluster derived.

From spectroscopic observations, one can also simultaneously derive the 
 individual masses of the stars and with complete spectra for all cluster 
 members, an independent and more direct determination of the IMF results. 
However because of spectroscopic sensitivity limits, the determination of 
masses is usually only possible for the bright stellar population. 
Since the monochromatic K magnitude of the cluster members can be
acquired for stars much fainter than the limit of spectroscopic methods,
the analysis of the NIR luminosity function is a particularly powerful tool
for investigating the IMF of faint stars in distant clusters
or stars at and below the hydrogen burning limit in nearby clusters.
Determining the fraction of cluster members at and below the
hydrogen burning limit is a holy grail of present stellar research.
The application of the luminosity function method to a nearby populous
cluster would provide a first glimpse into the brown dwarf population
formed at the time of a typical open cluster's birth.

\subsection{Comparing Models with Observations: The Trapezium Cluster}
\label{discusstrap}

The Trapezium cluster is a excellent system for evaluating
 the KLF modeling techniques developed in this paper.
It is the most densely populated and best studied nearby
 (D $\sim$ 400-450pc) cluster, and the central part of a much
 larger cluster known as the Orion Nebula Cluster (ONC).
The ONC has recently been studied by \cite{hil97},
who used optical spectroscopy to obtain a mean age for the cluster of 
$0.8\:\times\:10^{6}$ years and to construct an IMF for stars 
with masses primarily in excess of the hydrogen burning limit (HBL).
In addition, infrared imaging surveys have been made of both the
Trapezium cluster \citep{zin93,mcc95} and the ONC \citep{ali95}
enabling the construction of the cluster KLF. 
For comparison with our models, we consider only
the KLF for the Trapezium cluster, the 5' by 5' central core
of the ONC. 

We constructed a KLF of the Trapezium by combining the published KLFs 
 of \cite{zin93} and \cite{mcc95}.
Our adopted KLF for the Trapezium is shown in the top panel 
 of Figure \ref{trapfit}.
The Zinnecker et al. KLF includes the bright stars but is not
 complete at and below the HBL. 
The McCaughrean et al. KLF extends to very faint magnitudes, well 
 below the HBL for a one million year old cluster, 400 pc distant, 
 but because of source saturation, is incomplete for and does 
 not include bright stars.
Neither of these referenced cluster KLFs were corrected for contamination
 by foreground or background field stars.
In addition, neither was corrected for the effects of
 nebular contamination which would confuse the completeness
 of the surveys.
However, we compared this combined Trapezium KLF to the KLF
 from the \cite{ali95} survey of the ONC and found 
 good agreement in the location of the turnover, bright and
 faint ends of the two KLFs, although the Ali \& Depoy survey 
 was not as sensitive as that represented by the McCaughrean et al. KLF.  
We reiterate that the extent to which the combined KLF represents
 the true Trapezium KLF is uncertain because we cannot
 account for field star or nebular contamination.

Here our goal is to find the simplest functional from of an underlying
IMF whose resulting model KLF will best fit the observed KLF.
We constrained the star forming history of the Trapezium cluster
by using the mean age from \cite{hil97} i.e., 0.8 million years. 
We allowed an age spread of 1.2 million years ($\Delta\tau/\tau$ = 1.5) 
about this mean age, corresponding to constant star formation from 
0.2 to 1.4 million years ago.
We inspected the observed KLF and determined that a single power law
IMF could not satisfy the observations since the KLF has a peak
and turnover well above the completeness limits of the two surveys.
Therefore we began with a simple 2 power law IMFs.
We next used a three power law IMF with a flat (zero slope)
IMF in the middle. 
For symmetry, the two outer power-law slopes
were set to have equal but opposite sign slopes. 
We varied these outer slopes to have absolute values between 0.25 and 2.00 
and adjusted the mass range over which the middle slope of the IMF was flat.
Finally as a third set of experiments, we allowed the slope
of the middle power law to vary, still holding the outer
two slopes to have equal but opposite sign slopes.

We produced a suite of model KLFs for these different IMFs and
compared them to the combined Trapezium KLF using a chi-square
fitting procedure.
Simply, we normalized model KLFs to the observed KLF such
 that the model and observed KLFs contain the same number of stars
 between absolute K magnitudes of 0 and 6.5.
We then calculated the $\chi^2$ statistic and probability over 
 this K magnitude range.  
To derive a best fit, we compared a suite of model KLFs varying a single
 IMF parameter, e.g, the middle slope or one of the $m_{break}$ values
 and then determining the $\chi^2$ minima for that variable.
Model KLFs were created for a range of possible IMF parameters 
 and compared to the Trapezium KLF in this way.

Best fit model IMFs for each of the tested functional forms of the IMF
are listed in Table \ref{fittable}. 
Two power law fits in general were not good. 
Symmetric flat topped IMFs fit better and finally a slightly 
 rising IMF across the middle provided a best fit with $\chi^2\:\:\sim\:\:1$. 
Some variation in each of the parameters still allowed for a fit
of $\chi^2\:\:\sim\:\:1$ and examples are listed in Table \ref{fittable}.
The IMFs f and g produced best fits to the data:
For purposes of discussion, we adopt IMF (g) as representative of the 
Trapezium IMF and repeat its parameters here:
\begin{equation} \Gamma = \left\{ \begin{array} {l@{\quad:\quad}l}
 +1.35 &\:0.08\:M_\odot\:>\:M_\star  \\
 -0.25 &\:0.80\:M_\odot\:>\:M_\star\:>\:0.08\:M_\odot \\ 
 -1.35 &\:M_\star\:>\:0.80\:M_\odot
  \end{array} \right. 
\label{bestfit}
\end{equation}
The model KLF corresponding to IMF (g) is shown in the top panel of
 Figure \ref{trapfit} compared to the combined Trapezium KLF and compared
 to a model KLF calculated with the single power law slope \cite{sal55}
 field star IMF over the entire standard mass range.

From our modeling of the observed KLF for the Trapezium cluster 
we find that the predicted IMF has a rising slope
for intermediate mass stars, flattens around a solar mass,
reaches a peak near the HBL and turns over below the hydrogen burning limit.
There are several comparisons between the observed and modeled
Trapezium KLF and between the ONC IMF derived by Hillenbrand 
and our derived IMF (g) which should be made.
First, there exists a significant "tail" to the observed Trapezium
luminosity function which is not accounted for in the model KLFs.
No attempt was made to account for these very faint stars as cluster members
because if they were, they would require ages much older than
the distribution suggested by the HR diagram.
We instead suggest that these are in fact either extremely
embedded cluster members or heavily extincted background field stars
(A$_V$ $>$ 20 - 30). 
Either of these possibilities suggests
that our derived IMF is in fact an upper limit to actual IMF below 
the hydrogen burning limit.
Experiments studying the effects of extinction 
on the model KLF by \cite{mea96} and \cite{com96} found that while
extinction tended to shift the LFs to fainter magnitudes, the slope
of the KLF was preserved. 
Thus, the steeply falling slope at the low mass end of the derived 
Trapezium IMF is reflective of the actual underlying IMF.
However the true IMF may turnover at a larger mass than that
implied by our present models.
In the lower panel of Figure \ref{trapfit}, the mass function
derived from the Trapezium KLF is compared to that of \cite{hil97}
derived from spectroscopic observations.
The two mass functions are generally very similar.
In particular, these two mass functions agree very well at the 
 high mass end ($M_\star\;>\;2.0\;M_\odot$). 
For masses in the range 
 $2.0\;M_\odot\;>\;M_\star\;>\;0.5\;M_\odot$
 the IMF derived from modeling the luminosity function contains 
 more stars than that derived by Hillenbrand. 
However it is not clear how significant this difference is given
the possible systematic uncertainties involved in both methods for
determining the IMF and the fact that these two IMFs sample
different volumes of the Orion Nebula region.
For masses below $M_\star\;<\;0.1\;M_\odot$, the IMF
derived from the KLF modeling also contains considerably more
stars than the spectroscopic IMF. 
However this difference is also not likely to be significant either
since the spectroscopic IMF of \cite{hil97} is not complete
below 0.1 $M_\odot$.

Lastly, we can investigate whether the field star IMF (FSIMF) could 
also produce a KLF which reasonably matched the Trapezium KLF.
To test this, we used the recent field star IMF parameterization from
\cite{sca98}.
\cite{sca98} suggested a multiple power law IMF with the form:
\begin{equation}
 \Gamma = \left\{ \begin{array} {l@{\quad:\quad}l}
 -0.20  & 1.0\:M_\odot\:>\:M_\star\:>\:0.1\: M_\odot \\
 -1.70  & 10.\:M_\odot\:>\:M_\star\:>\:1.0\: M_\odot \\
 -1.30  & \:M_\star\:>\:10.\:M_\odot 
  \end{array} \right. 
\label{scaloimf}
\end{equation}
Comparing the IMF in equation \ref{bestfit} to the field star IMF 
 in equation \ref{scaloimf}, one finds that these two IMFs are quite 
 similar, although for stars in the range of $10.0\;>\;M/M_\odot\;>\;1.0$,
 the Scalo IMF is steeper than the IMF in equation \ref{bestfit}.
In addition the Scalo IMF does not extend below the hydrogen burning limit.
To facilitate comparison to the Trapezium data, we added a fourth
 power-law to the Scalo IMF to account for the faintest stars.
We varied $m_{break}$, the mass at which the fourth power-law begins,
 between 0.06 and 0.1 M$_\odot$.
In addition, we varied the slope of the fourth power-law 
 (between -1.0 and +2.0).
The best fits are also listed in Table \ref{fittable}. 
Using this modified field star IMF did yield a $\chi^2\:\:\sim\:\:1$ 
 with an IMF that breaks near the hydrogen burning limit and falls
 with a similar steep slope as in the prior IMF fits.

To the extent that our adopted KLF represents the true KLF of the cluster,
 our modeling suggests that the IMF for brown dwarfs in the Trapezium
 cluster falls relatively steeply with decreasing mass.
However because contamination due to reddened background stars and 
 incompleteness due to nebular confusion has not properly been taken into
 account in the construction of the Trapezium KLF, the form of
 the derived IMF below the hydrogen burning limit should be regarded
 with appropriate caution. 
As shown in \cite{ll95} and \cite{lad96}, one can use
 control-field observations (which are not available for this dataset) 
 to gauge the completeness and membership at the faint end of the LF.
Also, our present modeling has not included the effects of extinction
 and infrared excess. 
\cite{hil98b}, using the (I-K) diagnostic, found an average K band excess 
 of 0.35 among identified optically visible cluster members.
This average excess is smaller then the bins we have used to construct
 the Trapezium KLF, and therefore should have only a minor effect.
Yet we can improve the derived IMF for the Trapezium cluster by
 constraining these effects using multi-wavelength observations and
 infrared color-color diagrams of the cluster and of near-by control
 fields.
In a subsequent paper, we will use new JHK near-infrared observations
 and appropriate control fields to accurately determine
 the completeness of the Trapezium KLF and model the effects of extinction 
 and infrared excess on the cluster KLF and the derived IMF.

Overall, we conclude from our modeling that the IMF of the Trapezium
 cluster is well represented by a three power-law mass function
 with a high mass slope between -1.00 and -1.7, a break in slope between 1 
 and 0.6 M$_\odot$ followed by a relatively flat or slightly rising slope
 to the hydrogen burning limit and then falling with a steep slope 
 $\sim$ +1 through the brown dwarf regime.

\section{Conclusions}
\label{conc}

We have performed a series of experiments aimed at studying how the
 pre-main-sequence mass-to-luminosity relation, star forming history and 
 initial mass function each affect the form of the luminosity function 
 for populations of young pre-main sequence stars. 
Using models of the near-infrared luminosity function and varying these 
 primary inputs, we have derived the following simple conclusions 
 about model near-infrared luminosity functions:

\begin{itemize}
\item We find that the KLF of a young cluster is considerably more sensitive 
to variations in its underlying IMF than to either variations in the
star forming history or the PMS mass-to-luminosity relation.
\item PMS luminosity functions evolve in a systematic manner with increasing
 mean age and age spread. They evolve to fainter magnitudes and
 widen systematically with age.
\item The KLFs of young stellar populations are found to be generally
insensitive to variations in the adopted PMS mass-to-luminosity relations.
In the youngest, coeval clusters, the presence of deuterium burning can
produce significant features in the KLF which are sensitive to the 
adopted mass to luminosity relation. 
However even a small departure from a purely coeval star forming history will 
 render these features difficult to detect observationally.

\end{itemize}

We model the observed luminosity function of the Trapezium Cluster
 and are able to derive an underlying IMF for the cluster which spans
 a range of stellar mass from 5 $M_\odot$ to 0.02 $M_\odot$, well
 into the brown dwarf regime. 
The IMF we derive is the simplest multiple power-law function which can
 reproduce the observed luminosity function of the cluster given the
 mean age and star forming history derived from previous optical 
 spectroscopic studies \citep{hil97}.
The derived IMF for the Trapezium cluster consists of three power
 law segments, has a peak near the hydrogen burning limit and
 steadily decreases below the hydrogen burning limit and throughout
 the brown dwarf regime.
We derive a brown dwarf mass spectrum of the form dN/dlogm 
 $\sim\;m^{+1}$ (0.08$\;>\;M/M_\odot\;>\;$0.02).
However, the form of the IMF below the hydrogen burning limit
 must be regarded with caution since the faint end of the observed 
 cluster KLF has not been adjusted for the possible effects of background 
 star and nebular contamination.
Above the hydrogen burning limit, the Trapezium IMF is also 
 consistent with that recently advocated for field stars by \cite{sca98}.

\acknowledgments

We thank Richard Elston for conversations and suggestions on the
 use of Monte Carlo techniques. 
We also thank Jo\~{a}o Alves, Alyssa Goodman, Pavel Kroupa, John Scalo and John
 Stauffer for conversations and suggestions which improved upon this work.
A.A.M. is supported by a Smithsonian Predoctoral Fellowship. 
EAL acknowledges support from a Research Coorporation Innovation Award, a
 NSF PECASE Grant, AST-9733367, and a NASA ADP grant NAG5-6751 to the University
 of Florida, Gainesville.

\appendix

\section{Bolometric Corrections}
\label{boloappendix}

Bolometric corrections were interpolated from a table of basic
 stellar properties to convert model stars' luminosities and effective 
 temperatures into monochromatic magnitudes.
The following formulae were used to convert from these
 theoretical quantities to the monochromatic pass band magnitude.
\begin{eqnarray}
\mbox{M}_\lambda & = & \mbox{M}_{bol,\star} - \mbox{BC}_\lambda
\label{bc1} \\
\mbox{M}_{bol,\star} & = & \mbox{M}_{bol,\odot} - 2.5*\log(L/L_\odot)
\label{bc2} \\
\mbox{M}_\lambda & = & \mbox{M}_{bol,\odot} - 2.5*\log(L/L_\odot) - \mbox{BC}_\lambda 
\label{bc3} 
\end{eqnarray}
We have adopted the near-infrared colors of dwarf stars.
While the temperature scales for young pre-main sequence stars fall 
 somewhere between the temperature scales for main sequence dwarfs and 
 giants \citep{luh99}, their near-infrared colors are more dwarf-like
 \citep{luh99}.

We constructed our table of stellar properties beginning with those
 compiled in \cite{kh95} (KH95). 
We adjusted this tabulation to reflect the large temperature range of
 our model stars and to investigate some of the dependencies of our models
 on this tabulation.
For spectral types O3 to B0.5, corresponding to $\mbox{T}_{eff}$ from
30,000 to 50,000 K,  we used V band bolometric
 corrections and effective temperatures from \cite{vac96}.
O star colors were assumed degenerate at all near-infrared bands and assigned
 the colors of \cite{joh66}.
B star red and near-infrared colors were taken from a recent 
 tabulation by \cite{win97}. 
There is presently a significant study in the literature of the
 color-$\mbox{T}_{eff}$-spectral type relation for cool
 stars with $\mbox{T}_{eff}$ less than 3500K \citep{leg96}.
Because we related the stellar effective temperature directly to the stellar
 color and bolometric correction and do not assign spectral types to our
 model stars, we do not need to define any particular spectral sequence.
For the color-$\mbox{T}_{eff}$-$\mbox{BC}_{V}$ relation of M dwarfs,
 we used the relations compiled by Bessell et al. (1991,1995,1998).
However, our model stars have effective temperatures much cooler than
 those defined by the observationally defined scales.
To extend our color relations to even cooler effective temperatures,
 we used the broad band fluxes from model atmosphere 
 and PMS evolutionary calculations by \cite{bar98}.
Though typical model atmosphere calculations have been used to fit spectra 
 and develop spectral sequences for sometime, the predicted broadband 
 fluxes from these model atmospheres have been largely
 inaccurate \citep{leg92,grah92,kir93a,tin93,leg96}.
\cite{bar98} models are the most recent calculations 
 of both synthetic spectra and broadband fluxes for cool stars and 
 brown dwarfs and employ the most recent opacity tables.
After examining their different calculations and predicted
 cool stars colors, we used their 10 Gyr models to extend our
 color-$\mbox{T}_{eff}$-$\mbox{BC}_{V}$ tabulation at all
 red and near-infrared bands.
 
We checked our tabulation against other recent compilations of stellar
 properties and cool stars observations in the literature.
We compared our $\mbox{T}_{eff}$-$\mbox{BC}_{V}$ relation to 
those polynomial fits recently derived by \cite{flo96} and \cite{hil97}.
We found that our tabulation was in systematic disagreement with these fits.
The cause was traced to the original \cite{sk82}
 $\mbox{T}_{eff}$-$\mbox{BC}_{V}$ tabulation used in the KH95 compilation.
We refer to the discussion in \cite{bes98} (their Appendix D) 
 as to the source of this discrepancy and follow their prescription to
 add +0.12 to the \cite{sk82} $\mbox{BC}_{V}$ scale.
Combined with our choice of $\mbox{M}_{bol,\odot}$ equal to 4.75, this
 yielded a solar $\mbox{BC}_{V}$ equal to -0.07 and an absolute
 $\mbox{M}_{V,\odot}$ magnitude of 4.81.
We then smoothed the $\mbox{T}_{eff}$-$\mbox{BC}_{V}$
 relation to match those of \cite{hil97} and \cite{flo96}.

A more important concern is the accuracy of our tabulation below 3500K.
We tested our tabulation for cool stars by compiling, from the literature,
 observed colors, effective temperature determinations and bolometric 
 luminosities for M dwarfs from the literature including 
 published data by \cite{ber92}, \cite{tin93}, 
 \cite{jon94} and \cite{leg96}.
We used repeat observations and derivations of stellar properties
 (e.g. $\mbox{T}_{eff}$) of the same M dwarfs but by different authors
 as reflecting different spectral type to effective temperature
 scales and variability among late-type stars, as well as fundamental
 uncertainties in the computation of these values.
The observed M dwarf colors, bolometric corrections, and 
 effective temperatures were compared to our compiled table.
First, our comparison indicated that the compiled bolometric 
 corrections were consistent with the observed M dwarf values.
Second, and more importantly, our comparison showed that for
 the near-infrared bands, specifically the K band, the bolometric 
 corrections are insensitive to effective temperature scales 
 for low mass stars and brown dwarfs.

\section{Combining Pre-Main Sequence Tracks}
\label{pmsappendix}

A set of theoretical evolutionary sequences were used to convert the masses
and ages of model stars to luminosities and effective temperatures
which were then converted to monochromatic passband magnitudes using the
color tabulation documented in Appendix \ref{boloappendix}.
These sequences were a combination of three different regimes:
a theoretical zero age main sequence, a set of intermediate
mass PMS tracks and a set of sub-solar and brown dwarf PMS
stellar models.
When converting the model stars, the algorithm determined if, 
for certain stellar age, the star would be found on the ZAMS 
or in a PMS stage. 
If the star would be found on the ZAMS, then it luminosity and effective
temperature was derived from the ZAMS.
If the star would be in a PMS stage, then the PMS models were used.

Unfortunately, PMS theoretical models are not typically calculated
for the entire mass range from brown dwarfs to high mass B stars. 
Because of this, we combined two different sets of PMS tracks to 
provide a complete mass range.
We took the opportunity to use different sets of PMS tracks for high 
and low mass stars to account for an apparent 
mass-age correlation found by many authors who have used
PMS tracks to derive real ages and masses for stars using the HR 
diagram \citep{hil94,mey96,hil97}.
These authors point out that when masses and ages are derived for a 
cluster of real stars using PMS models, a correlation existed such that
the more massive stars were systematically older than lower mass stars.
These authors suggested that the cause of this correlation is 
due to the way canonical PMS tracks are constructed.
Canonical PMS tracks evolve the model stars from infinite spheroids.
More recent studies suggest that stars evolve during a protostellar phase
along a specific mass-radius relationship referred to as the proto-stellar
birthline \citep{sta83,ps93}.
Using a proto-stellar birthline as the initial condition for PMS tracks 
will most prominently affect the predicted luminosities and effective
temperatures of the youngest and highest mass stars,
where the stars' proto-stellar (birthline) lifetimes are comparable with
these stars' pre-main sequence lifetimes.

Rather than using canonical PMS tracks for model stars with
 masses greater than solar, we used "accretion scenario" PMS model
 calculations by \cite{ps93} and \cite{ber96a}.
Accretion scenario PMS models better represent the location
 of the young intermediate mass stars on the HR diagrams
 \citep{ps93,ber96b}.
Yet the accretion scenario PMS tracks cannot be straightforwardly 
used with sub-solar mass canonical PMS tracks. 
We adopted the accretion scenario tracks for models above 2 solar masses and
canonical tracks below 1 solar mass. 
Between these two limits, we compared the canonical and accretion 
scenario calculations.
We performed an average of the accretion scenario and canonical mass 
tracks, weighting each average to result in a smooth conversion from 
the canonical (sub-solar mass) to accretion scenario (high) regimes.
We examined the theoretical HR diagram resulting from our combination of
ZAMS, accretion scenario, averaged, and canonical PMS tracks.
This new set of tracks and resulting isochrones were found to be smooth between
all regimes and was used as input for the luminosity function modeling.

\newpage

\newpage

\figcaption[figure1.eps]
{
Definitions of age and age spread used to define the cluster star forming
 history in modeling the cluster KLF throughout this paper.
The mean age, $\tau$, in this simple model is equivalent to the average 
 of the ages of the oldest and youngest stars assuming a constant rate 
 of star formation throughout the star forming history of the cluster.
\label{age_agespread}
}

\figcaption[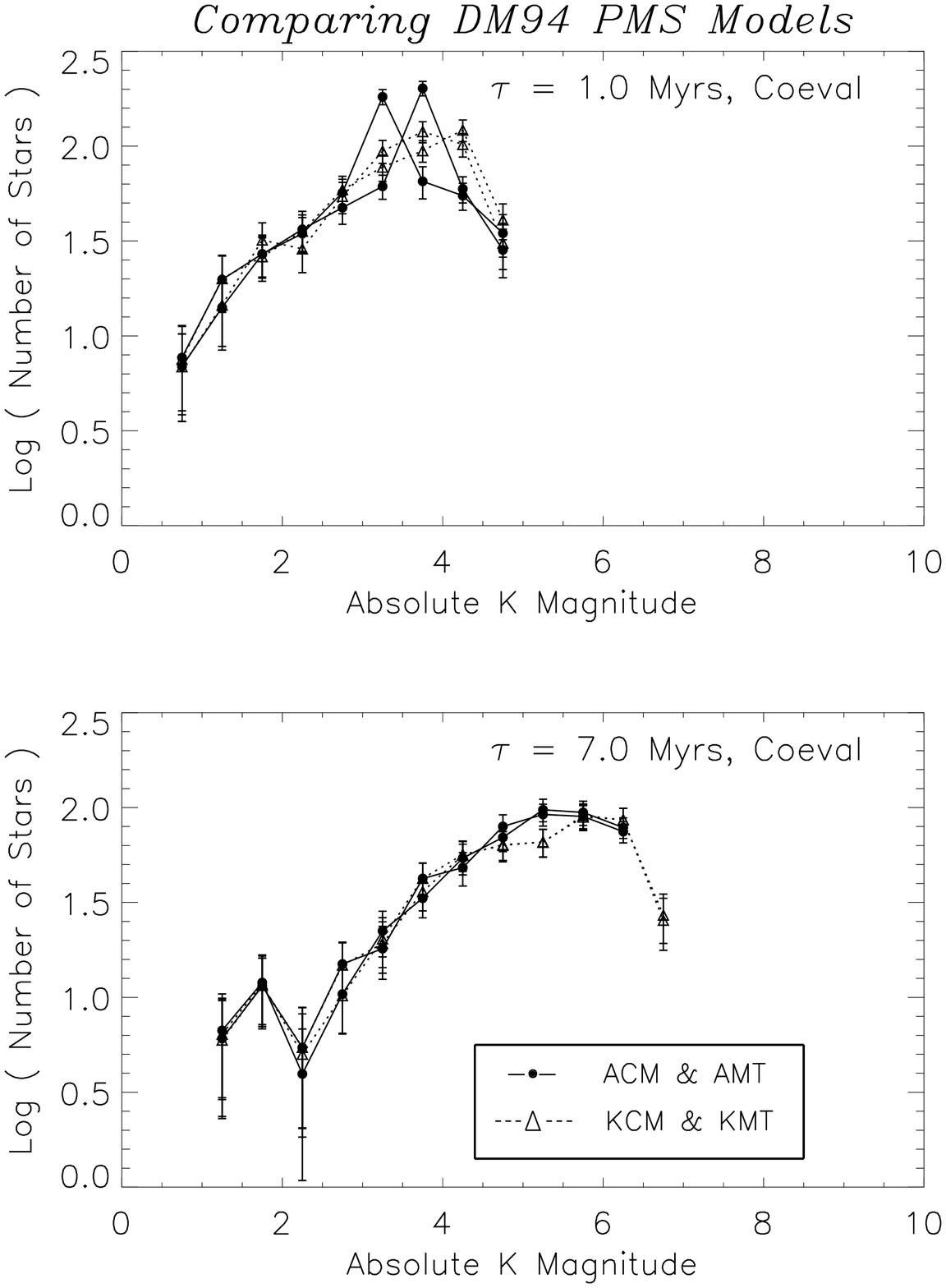]
{
Model KLFs derived from the four sets of PMS models of DM94. 
Each corresponds to a different combination of input physics 
 as described in DM94 (see also Table \ref{pmstable}). 
These model KLFs used a log-normal IMF with a lower mass limit of 0.1
 $M_\odot$ and coeval star formation with mean ages
 of 1 (top) and 7 (bottom) Myrs.
Different symbols correspond to different
 input opacity tables in the PMS models used.
The number of stars in each bin corresponds to the mean value 
 of that bin for 100 independent realizations of the model KLF.
Each realization of the model KLF contains 1000 stars. 
Error bars are 1 $\sigma$ standard deviation of the mean value of that bin
 for the 100 iterations of the model. 
\label{dm94compare}
}

\figcaption[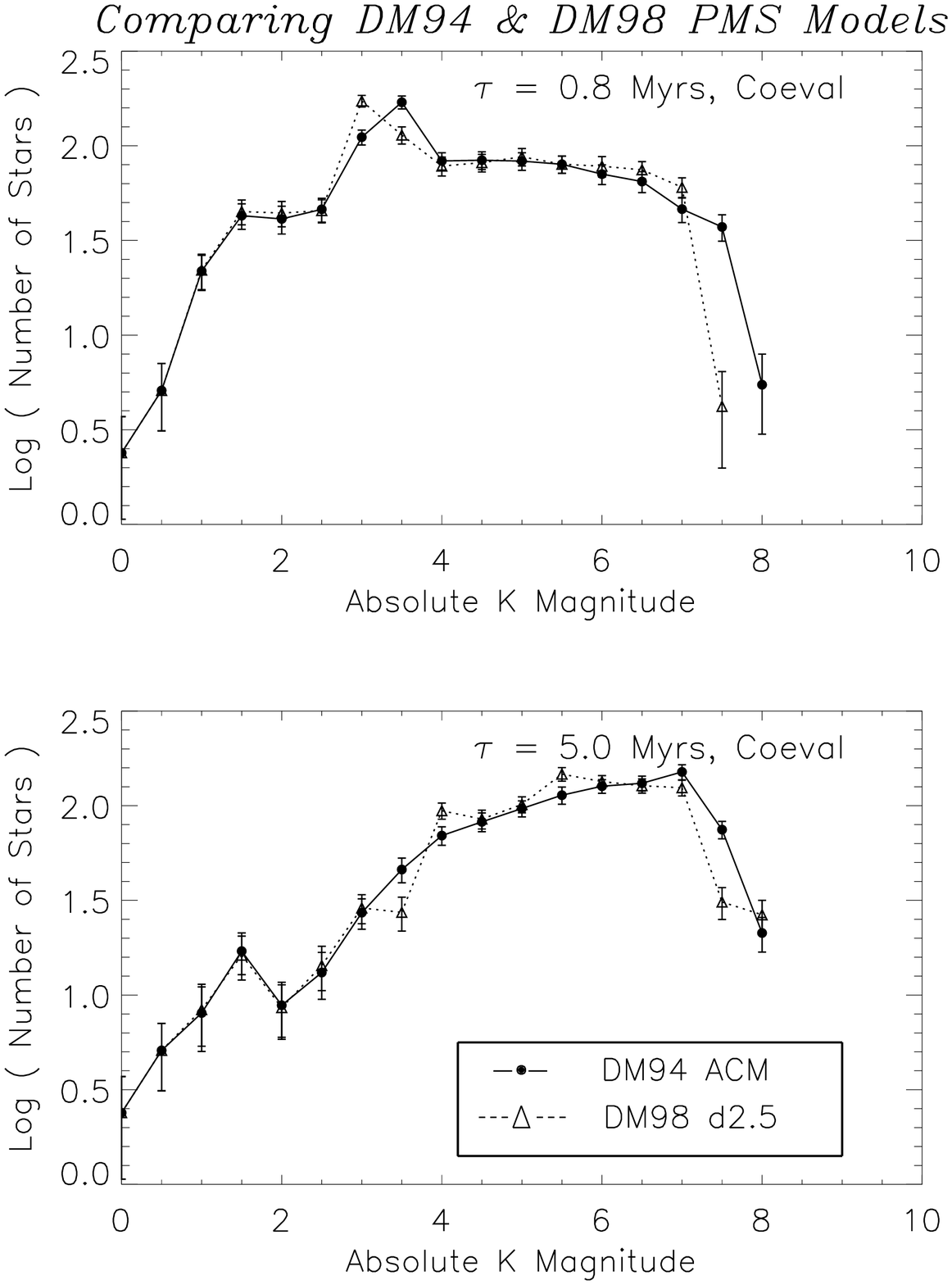]
{
Model KLFs comparing the input PMS models from DM94 and DM98.
Shown are model KLFs computed using the ACM model of DM94 and the d2.5 
 model from DM98 (see Table \ref{pmstable}).
These two PMS evolutionary models differ in basic input 
 physics such as opacity table, equation of state and treatment of convection.
However they have identical mass ranges and deuterium abundances.
Both panels correspond to model KLFs for clusters with a coeval 
star forming history and mean ages of 0.8 Myrs 
(top panel) and 5 Myrs (bottom panel). 
Error bars are the same as those in Figure \ref{dm94compare}. 
\label{dm94vdm98}
}

\figcaption[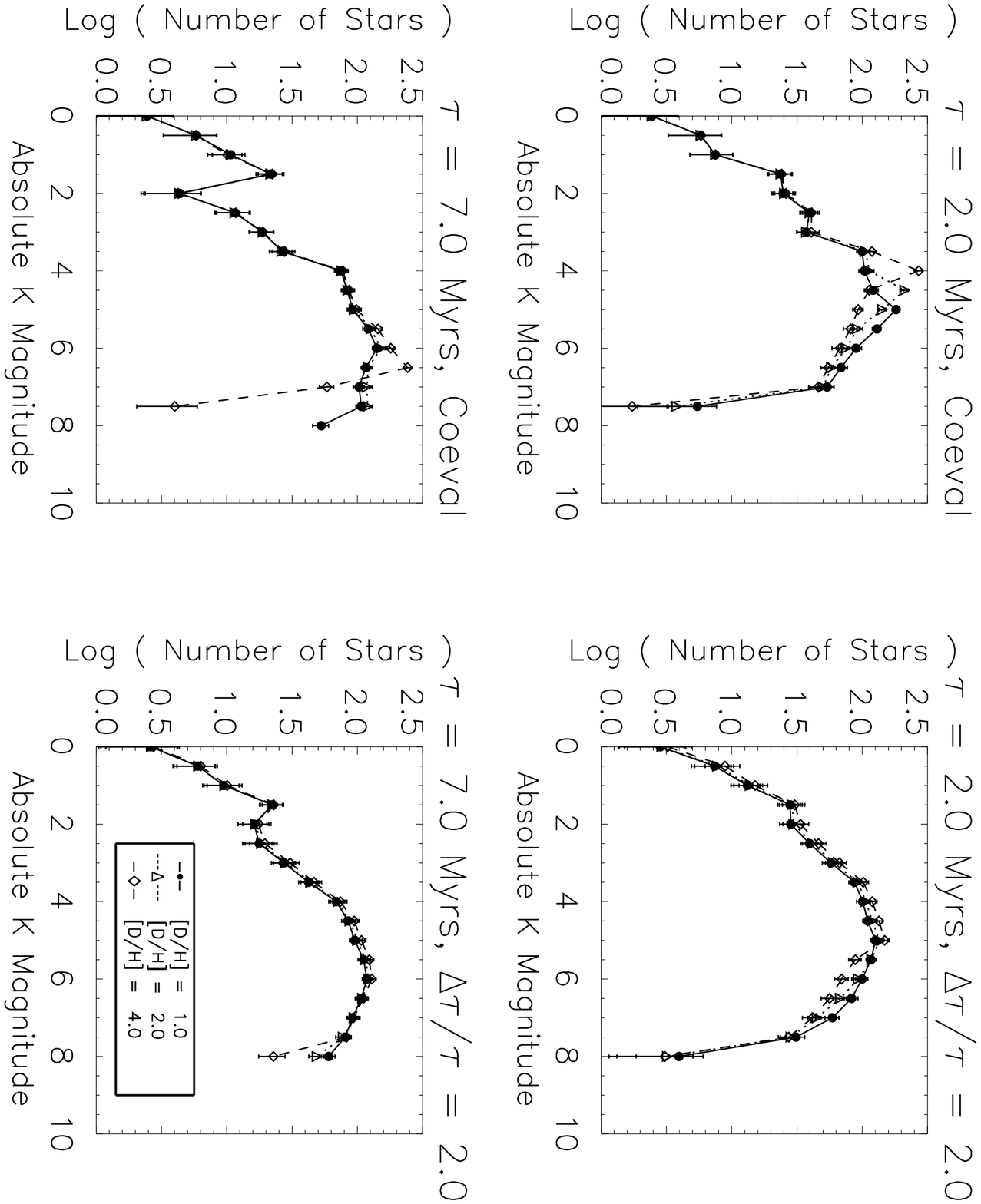]
{
Model KLFs derived from DM98 PMS models varying the fractional deuterium 
 abundance at the onset of pre-main sequence contraction. 
These model KLFs have the standard log-normal IMF sampled over the complete
mass range available for the DM98 PMS models (see Table \ref{pmstable}).
The first column of panels are for coeval SFHs with mean ages of 2 and 7 Myrs.
In the second column of panels, we display model KLFs with the same mean ages 
but with age spreads that are twice the value of the mean age.
The different model KLFs are labeled corresponding to the 
 deuterium abundance ratio relative to hydrogen
 and given in units of $\times\: 10^{-5}$.
Physically, these abundances represent one half, one and two times  
 the measured interstellar medium abundance of deuterium. 
Error bars are the same as those in Figure \ref{dm94compare}. 
\label{dm97compare}
}

\figcaption[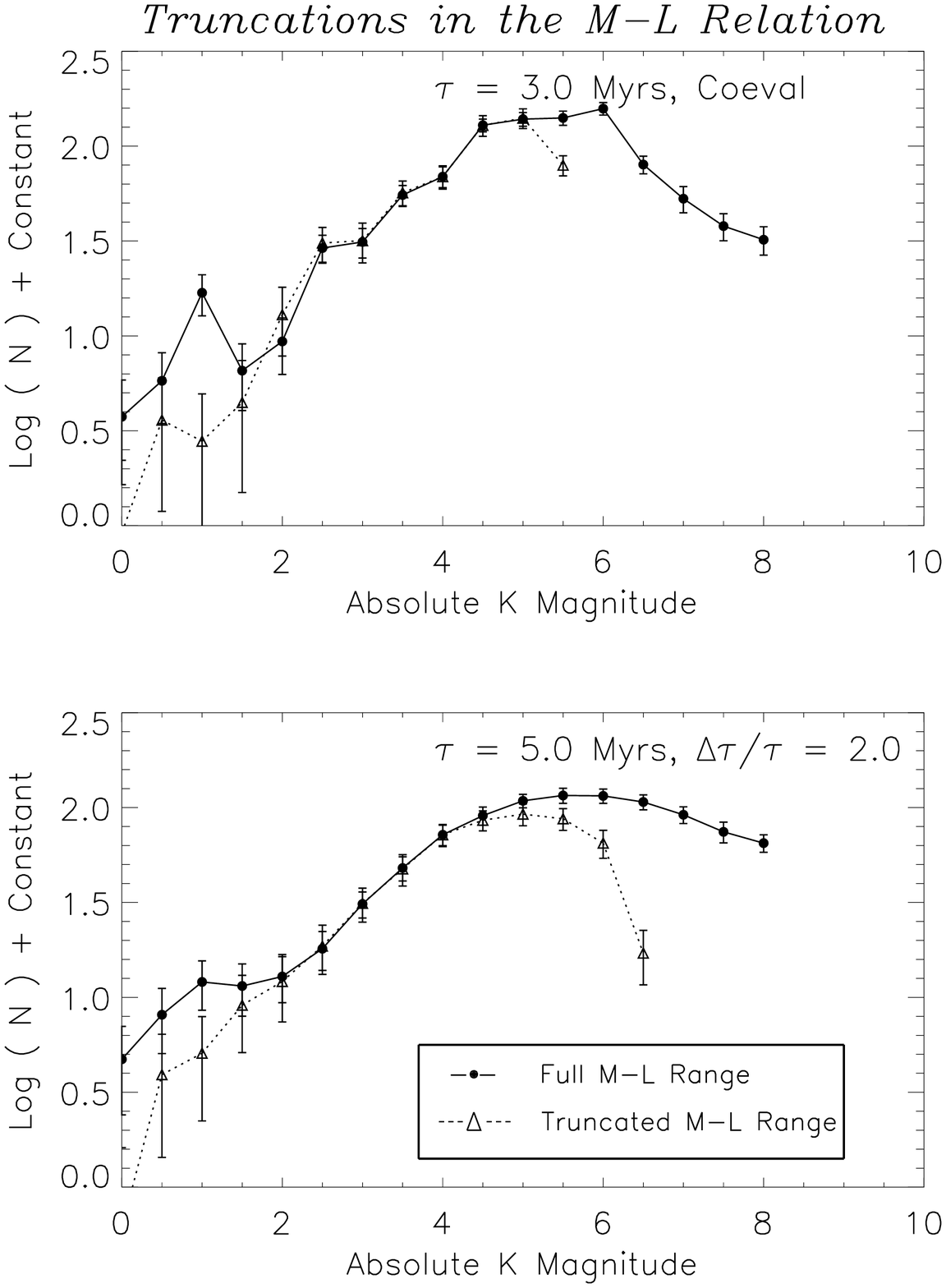]
{
Model KLFs testing the inclusion into model KLFs of high and intermediate mass
 stars as well as stars at the hydrogen burning limit and brown dwarfs.
The mass to luminosity relation was extracted from the ACM 
 PMS model of DM94, intermediate mass PMS tracks of \cite{ps93} and
 a ZAMS from \cite{sch92} for higher mass stars.
Two different mean ages and SFH histories are shown for illustration.
In the upper panel, model KLFs created using a coeval star forming history 
 and a mean age of 3.0 Myrs. 
The lower panel shows a model KLF with a continuous star forming
 history over the lifetime of the cluster and a mean age of 5 Myrs.
Error bars are the same as those in Figure \ref{dm94compare}.
\label{mltrunc}
}

\figcaption[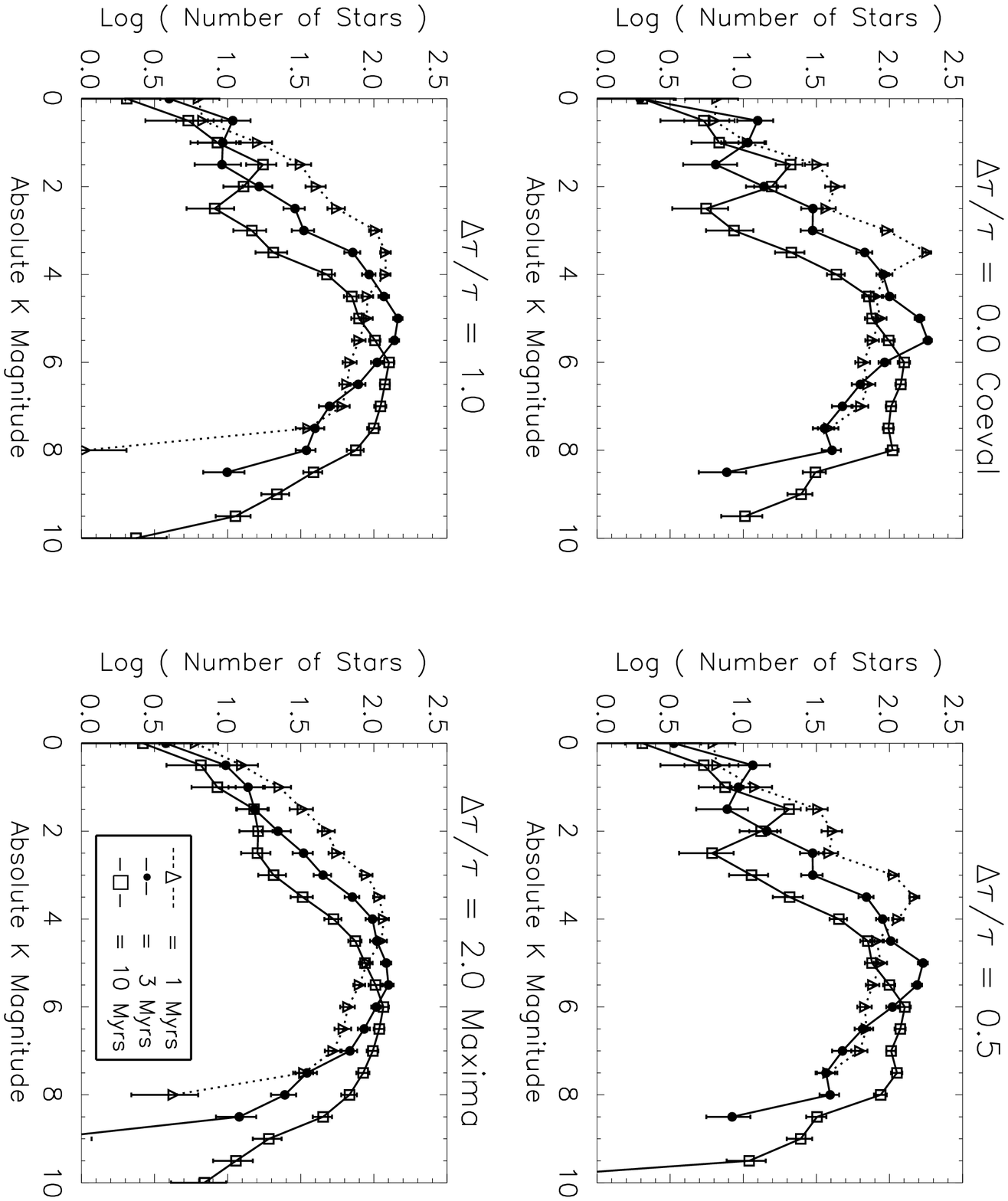]
{
Model KLFs varying the cluster mean age and age spread. 
Each panel displays a different $\Delta\tau/\tau$.
Mean ages shown are 1, 3 and 10 million years. 
Note that from panel to panel, features in the model KLF caused by inflections
in the M-L relation are smoothed by increased age spread.
The apparent break in the model KLFs in the last bin of 
 each KLF is primarily due to incompleteness in that bin due to a 
 truncation of the M-L relation at 0.02 $M/M_\odot$.
Please see section \ref{expimf} for further explanation.
Error bars are the same as those in Figure \ref{dm94compare}.  
\label{meanages}
}

\figcaption[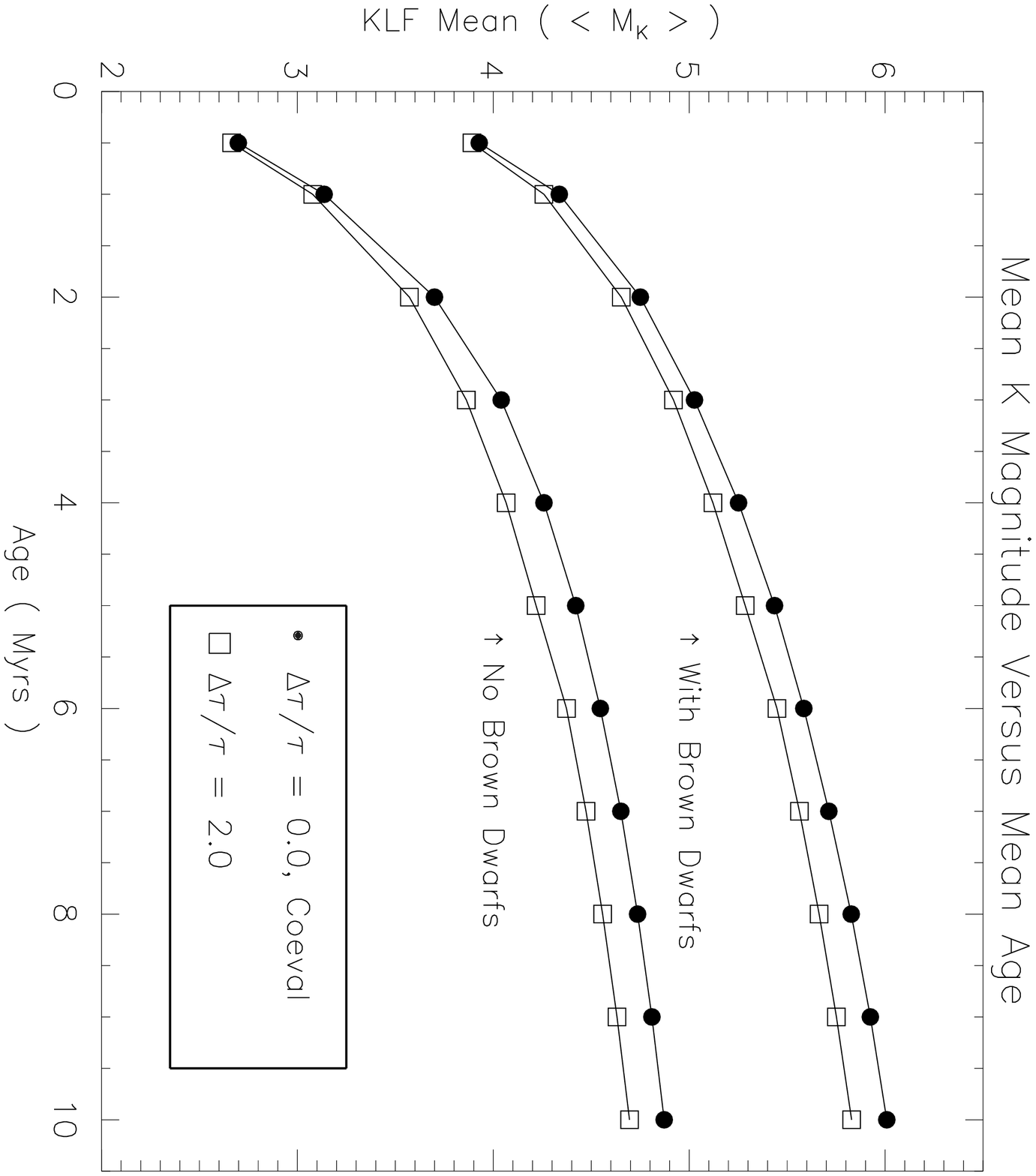]
{
Plot illustrating the evolution of the KLF mean with increasing cluster
 age and age spread.
The KLF mean refers to the arithmetic mean of the K magnitudes for
all synthetic cluster members.
Two sets of values are plotted for KLFs having two different underlying IMFs.
"With Brown Dwarfs" contains stars below 0.1 M$_\odot$
and "Without Brown Dwarfs" has no stars less than 0.1 M$_\odot$.
For each set of curves, the KLF mean was plotted for the two
extremes of possible star forming histories.
Error bars are not shown but are within the size of the plotting
symbols for a cluster of 1000 stars.
\label{meank}
}

\figcaption[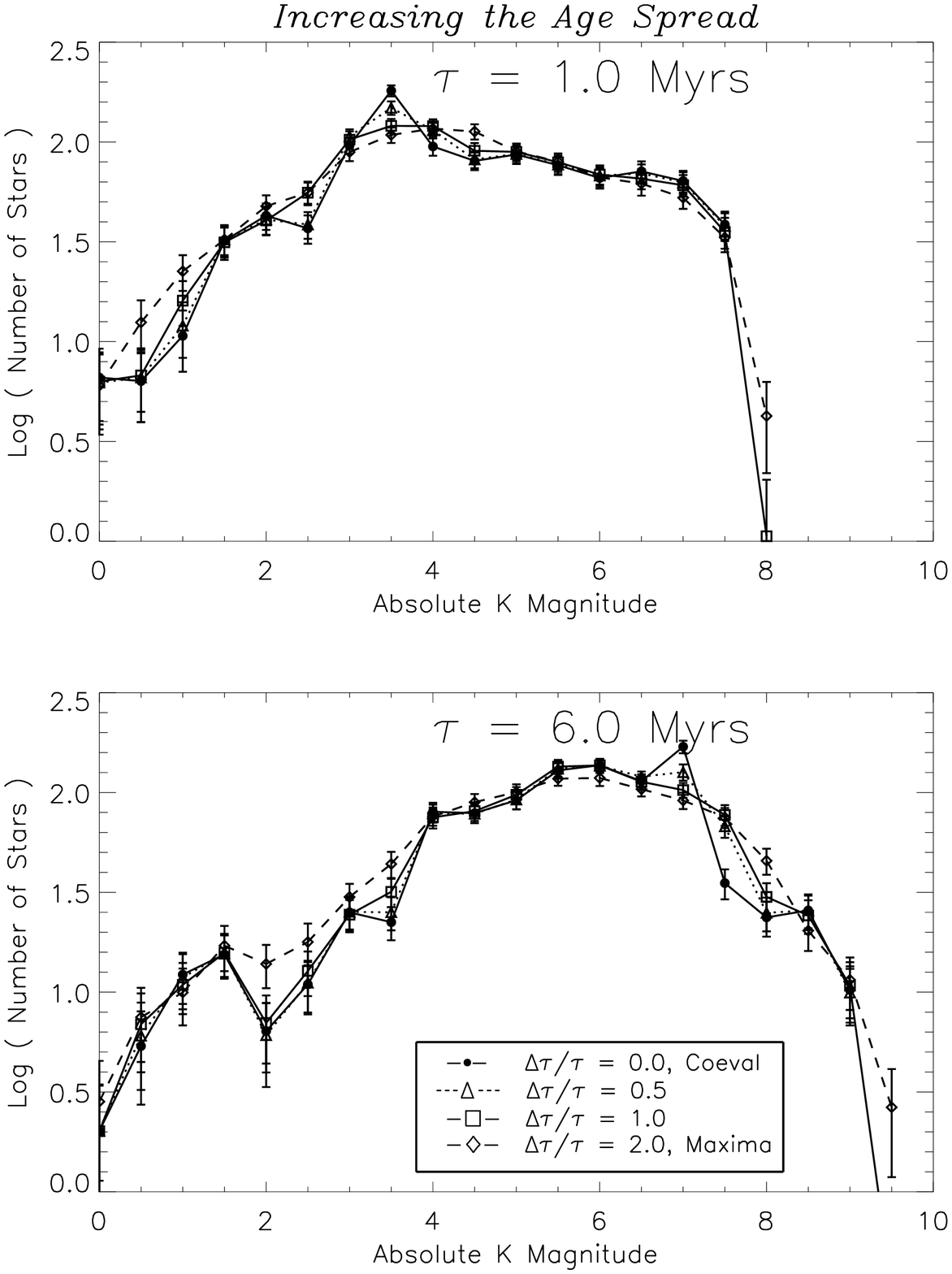]
{
Model KLFs plotted to show the effects of varying the cluster age spread.
Synthetic cluster KLFs with mean ages of $\tau = $ 1 and 6 Myrs,
 respectively, are shown in each panel. 
For each mean age (in each panel)
 the same four age spreads from Figure \ref{meanages} are over-plotted.
Increasing age spread erases features in the KLF caused by inflections 
 in the mass to luminosity relation.  
Error bars are the same as those in Figure \ref{dm94compare}.
\label{deltat}
}

\figcaption[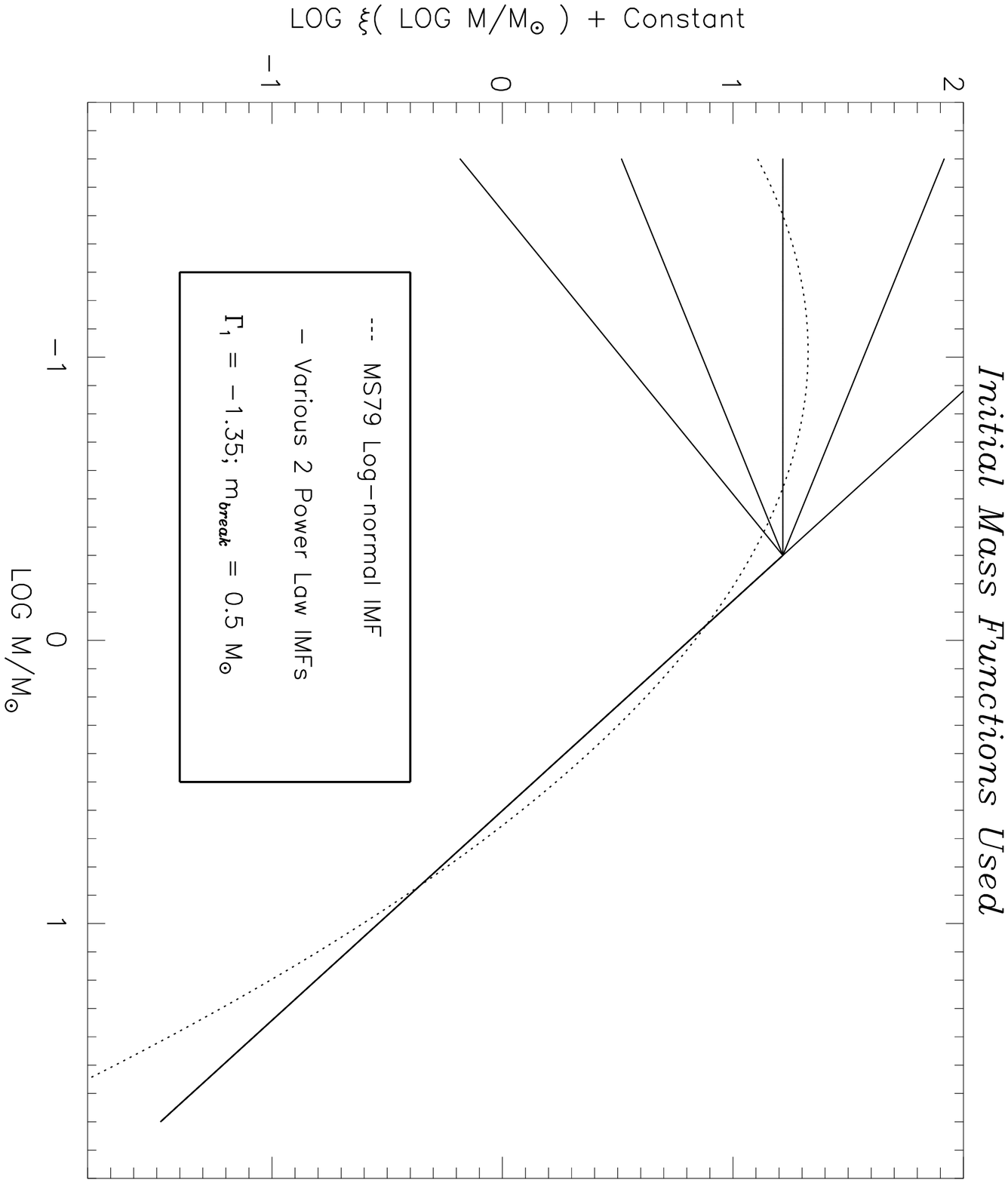]
{
Different mass functions used in modeling experiments.  
A few different parameterizations are shown. 
The log-normal form follows the parameterization of MS79 and
is extended to the mass limit, 0.02 $M_\odot$, of our model KLFs.
The other IMFs shown are examples of the simple 2 power law IMFs 
used for experiments in section \ref{expimf}. 
The high mass IMF slope, $\Gamma_{1}$, equals -1.35 equivalent
 to the power law IMF derived by Salpeter (1955). 
This power-law breaks at a mass, $m_{break}$, equal to 0.5 M$_\odot$.
Below the break mass, the IMF is governed by a low mass slope, $\Gamma_{2}$,
for which we show five different values, -1.35, -1.00, -0.40, 0.00, 
+0.40, and +1.0.
\label{massfunc}
}

\figcaption[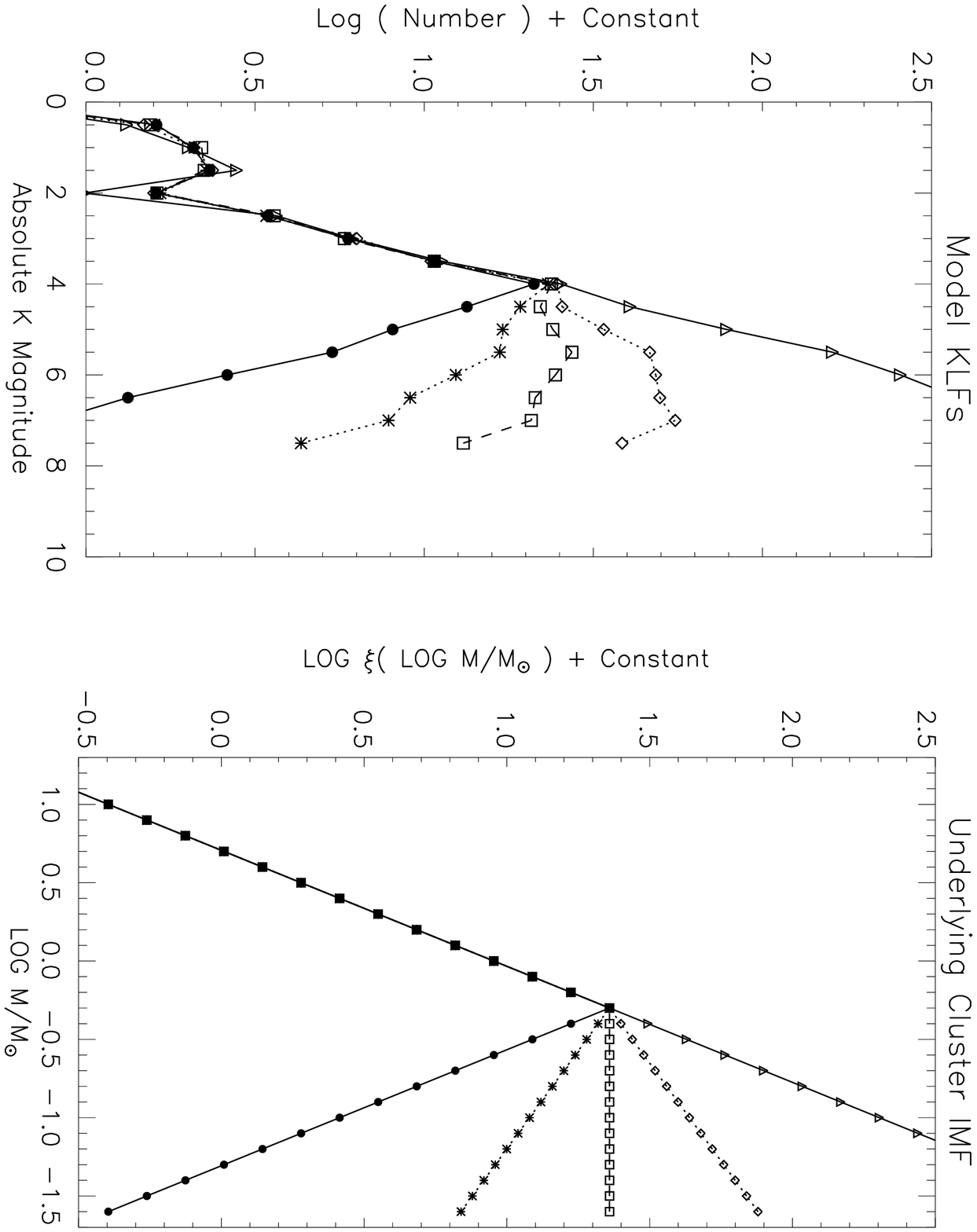]
{
KLF models using the 2 power law IMF described in equation \ref{2poweq}. 
The different model KLFs are normalized to have identical bright slope
 LFs where their underlying IMFs are identical.
The left panel shows the resulting KLFs corresponding to the underlying 
 IMFs shown in the right hand panel. 
Symbols are identical for underlying IMFs and the resulting model
KLF.
\label{2powerklf}
}

\figcaption[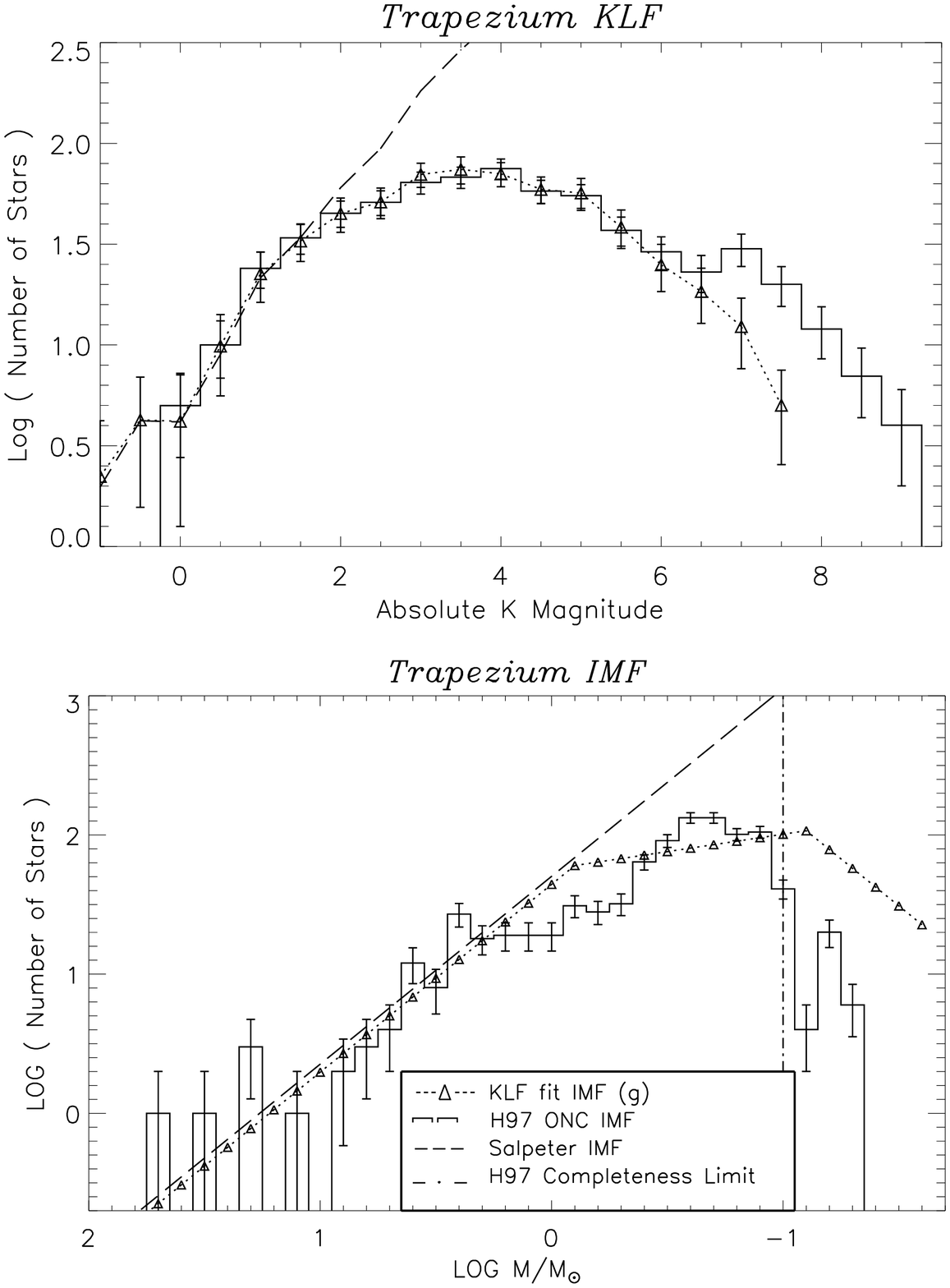]
{
Top panel: The histogram is the combined KLF for the Trapezium from \cite{zin93} 
 and \cite{mcc95} KLFs.
The model KLF is shown fit to the Trapezium KLF and uses the model (g) IMF
 from Table \ref{fittable}.
Also shown is a model cluster KLF created using the \cite{sal55} field
 star IMF extended to the mass limit (0.02 M$_\odot$).
The model SFH is characterized by a constant star formation rate, a mean age 
 of 0.8 Myrs \citep{hil97} and an age spread of 1.2 Myrs.
Lower panel: Histogram of the Orion Nebula Cluster (ONC) IMF derived 
 by \cite{hil97} compared to model IMF (g) used for the fit model KLF 
 in the upper panel.
Also shown are the \cite{sal55} field star IMF and the completeness
 limit in mass quoted by Hillenbrand (1997,1998).
For comparison, model IMF (g) is scaled to contain the same number of stars 
 as the ONC IMF above its completeness limit.
Error bars for the Hillenbrand IMF reflect one sigma counting statistics. 
The model KLF error bars are one sigma standard deviations 
 as described in Figure \ref{dm94compare}.
\label{trapfit}
}

\clearpage

\begin{deluxetable}{ccccc}
\small
\tablecaption{Pre-main Sequence Models Used \label{pmstable}}
\tablewidth{0pt}
\tablehead{
\colhead{Model Name} &\colhead{Opacity Table} &\colhead{Convection Model} &
\colhead{$[D/H]$\tablenotemark{a}}&\colhead{Mass Range (M/M$_\odot$)} 
}
\startdata
\cutinhead{\cite{dm94}}
ACM 	&\cite{alex89} &FST\tablenotemark{1}	&2.0 	&0.018 $\rightarrow$ 2.5 \\
AMT 	&\cite{alex89} &MLT\tablenotemark{2}  	&2.0 	&0.1 $\rightarrow$ 2.5 	 \\
KCM 	&\cite{kur91}  &FST\tablenotemark{1}	&2.0 	&0.1 $\rightarrow$ 2.5 	 \\
KMT 	&\cite{kur91}  &MLT\tablenotemark{3} 	&2.0 	&0.1 $\rightarrow$ 2.5 	 \\[0.1in]
\cutinhead{\cite{dm98}\tablenotemark{b}}
d1.5 	&\cite{alex94} 	&FST\tablenotemark{4} 	&1.0 	& 0.017 $\rightarrow$ 1.5 \\
d2.5 	&\cite{alex94}	&FST\tablenotemark{4} 	&2.0 	& 0.017 $\rightarrow$ 3.0 \\
d4.5 	&\cite{alex94}	&FST\tablenotemark{4} 	&4.0 	& 0.017 $\rightarrow$ 3.0 \\
\enddata

\tablenotetext{a}{Deuterium Abundance relative to Hydrogen; 
In units of $\times 10^{-5}$}
\tablenotetext{b}{DM98 models were initially released in 1997.  These models were updated in 1998. The model used were those of the updated calculations.}
\tablenotetext{1}{Full Spectrum Turbulence Model; \cite{can90,can92}}
\tablenotetext{2}{Mixing Length Theory; $1/H_p = 1.2$}
\tablenotetext{3}{Mixing Length Theory; $1/H_p = 1.5$}
\tablenotetext{4}{Full Spectrum Turbulence Model; \cite{can96}}

\end{deluxetable}

\clearpage

\begin{deluxetable}{ccccccc}
\small 
\tablecaption{Model IMF Fits to Trapezium KLF\label{fittable}}
\tablewidth{0pt}
\tablehead{
\colhead{Name} &\colhead{$\chi^2$ Prob.} &\colhead{$\Gamma_{0}$} &
\colhead{$m_{break}$} &
\colhead{$\Gamma_{1}$} &\colhead{$m_{break}$} &\colhead{$\Gamma_{2}$}
}
\startdata
\cutinhead{Two Power Law Fits}
a &0.38 &- &- &-0.50 &0.10 &+1.00  \\[0.1in]
\cutinhead{Three Power Law Fits}
b &0.71 &-0.75 &0.25 & 0.00 &0.10 &+0.75  \\ 
c &0.86 &-1.00 &0.40 & 0.00 &0.08 &+1.00  \\ 
d &0.88 &-1.00 &0.60 &-0.25 &0.10 &+1.00  \\ 
e &0.93 &-0.75 &0.25 &-0.25 &0.10 &+0.75  \\ 
f &0.99 &-1.00 &0.70 &-0.25 &0.08 &+1.00  \\ 
g &0.99 &-1.35 &0.80 &-0.25 &0.08 &+1.35  \\
\cutinhead{Scalo 1998 Field Star IMF\tablenotemark{1}  + BDIMF}
h &0.96 &-1.70 &1.00 &-0.20 &0.10 &+0.75  \\
i &0.99 &-1.70 &1.00 &-0.20 &0.08 &+1.00  \\
\enddata
\tablenotetext{1}{Above 10 $M_\odot$, this IMF has a $\Gamma$ equal to -1.30.}
\end{deluxetable}
\clearpage
\pagestyle{myheadings} \markright{Figure}
\pagenumbering{arabic}
\setcounter{page}{1}
\epsscale{.9}
\plotone{figure1.eps}
\clearpage
\plotone{figure2.eps}
\clearpage
\plotone{figure3.eps}
\clearpage
\plotone{figure4.eps}
\clearpage
\plotone{figure5.eps}
\clearpage
\plotone{figure6.eps}
\clearpage
\plotone{figure7.eps}
\clearpage
\plotone{figure8.eps}
\clearpage
\plotone{figure9.eps}
\clearpage
\plotone{figure10.eps}
\clearpage
\plotone{figure11.eps}

\end{document}